\documentclass[pra,twocolumn,groupedaddress,showpacs,preprintnumbers,amsmath,amssymb]{revtex4}
\usepackage{graphicx}
\usepackage{dcolumn}
\usepackage{bm}

\newcommand{\abdiff}{$|qm-cl|_1$ }
\newcommand{\abdiffm}{|qm-cl|_1}
\newcommand{\diff}[2]{\frac{\partial #1}{\partial #2}}
\newcommand{\diffm}[3]{\frac{\partial^{#3} #1}{\partial #2^{#3}}}
\newcommand{\bra}[1]{\langle #1|}
\newcommand{\ket}[1]{|#1\rangle}

\newcommand{\avg}[1]{\langle #1 \rangle}
\newcommand{\eqcite}[1]{Eq. (\ref{#1})}
\begin{document}

\title{The Quantum Mechanics of Hyperion}

\author{N. Wiebe}
\affiliation{
Physics Department, Simon Fraser University, Burnaby, British Columbia, Canada V5A 1S6
}

\author{L. E. Ballentine}
\email{ballenti@sfu.ca}
\affiliation{
Physics Department, Simon Fraser University, Burnaby, British Columbia, Canada V5A 1S6
}

\begin{abstract}
This paper is motivated by the suggestion [W. Zurek, Physica Scripta, T76, 186 
(1998)] that the chaotic tumbling of the satellite Hyperion would become 
non-classical within 20 years, but for the effects of environmental decoherence.
The dynamics of quantum and classical probability distributions are compared 
for a satellite rotating perpendicular to its orbital plane, driven by the 
gravitational gradient.  The model is studied with and without environmental 
decoherence.  Without decoherence, the maximum quantum-classical (QC) 
differences in its average angular momentum scale as $\hbar^{2/3}$ for 
chaotic states, and as $\hbar^2$ for non-chaotic states, leading to 
negligible QC differences for a macroscopic object like Hyperion.  
The quantum probability distributions do not approach their 
classical limit smoothly, having an extremely fine oscillatory structure 
superimposed on the smooth classical background.  For a macroscopic object, 
this oscillatory structure is too fine to be resolved by any realistic 
measurement.  Either a small amount of smoothing (due to the finite resolution 
of the apparatus) or a very small amount of environmental decoherence is 
sufficient ensure the classical limit.  Under decoherence, the QC differences 
in the probability distributions scale as $(\hbar^2/D)^{1/6}$, where D is 
the momentum diffusion parameter.  We conclude that decoherence is not 
essential to explain the classical behavior of macroscopic bodies.
\end{abstract}


\maketitle
\section{Introduction}

Quantum mechanics (QM) is a more fundamental theory than is 
classical mechanics (CM).  This fact does not diminish the utility 
of CM in describing the behavior of macroscopic objects.  But 
theoretical consistency demands that the observed classical 
phenomena should also emerge from QM in an appropriate limit.  
However, a detailed understanding of this quantum-to-classical (QC) 
limit is remarkably difficult to achieve, and the origin of classical 
behavior is even subject to a degree of controversy.  There has been 
some confusion as to which QM structures should be compared with 
classical predictions, and hence to what the appropriate criterion for 
classicality should be.  The role that decoherence due to 
environmental perturbations might play in establishing the QC limit 
is also controversial.

A criterion for classicality that has often been used is based on 
Ehrenfest's theorem \cite{ehrenfest}.  If the 
size of the quantum state is small compared to the scale on which the 
potential energy varies, then the centroid of the state will 
approximately follow a classical Newtonian trajectory.  The time 
duration  (and range of other relevant parameters) within which this 
condition holds is referred to as the \textit{Ehrenfest regime}.  Since 
any wave packet will eventually spread until it reaches the size of 
the system (harmonic oscillators being a unique exception), it follows 
that this criterion for classicality will eventually fail, no matter 
how 
macroscopic the system may be.  If the \textit{break time} when this occurs 
were as long as the age of the solar system, there would be no cause 
for concern.  But in chaotic systems, the size of a wave packet grows 
exponentially, and the break time can be quite small.  A striking 
example, given by Zurek \cite{zurek2}, is the 
chaotic tumbling of Hyperion (a moon of Saturn), for which the 
Ehrenfest criterion for classicality will fail in less than 20 years.  
Zurek argues that \textit{environmental decoherence} can remove 
this paradox and restore classical behavior to Hyperion.  
This claim is certainly not correct, as long as the Ehrenfest criterion 
for classicality is used.  However, we propose another criterion for 
classicality, within which the role of decoherence will be reassessed.

The origin of the above paradox consists in a failing to take proper 
account of the statistical nature of QM.  It is not correct to identify 
the trajectory of a body with the motion of the centroid of the wave 
function.  QM does not describe the actual observed phenomenon, but 
only the probabilities of the various possible phenomena.  We should, 
therefore, compare quantum probabilities with classical probabilities.  
This leads us to define the \textit{Liouville regime} of quantum-
classical correspondence, within which the quantum probabilities are 
approximately equal to the classical probabilities that satisfy the 
Liouville equation.  There is no requirement that the probability 
distributions be narrow, and so the Liouville regime of classicality is 
usually much larger than the Ehrenfest regime.

The superiority of the Liouville criterion for classicality to the 
Ehrenfest criterion has been demonstrated in several ways.  
One way to see this is through the correction terms to Ehrenfest's theorem.  
When the width of the state is small but not negligible, these corrections 
can be obtained as a series involving the 
variance and higher moments of the position probability distribution 
\cite{balyang}.  
This series has exactly the same form for the classical Liouville 
probability distribution.  Hence Ehrenfest's theorem merely 
asserts that if the (quantum or classical) probability distribution is 
sufficiently narrow, its centroid will follow a Newtonian trajectory.  
To leading order, the deviations from the Newtonian trajectory are 
the same in both the quantum and classical cases, and so those 
deviations are not primarily quantal in origin.  The magnitude of the 
deviations from Ehrenfest's theorem depends primarily on the width 
of the initial state in configuration space, and has no 
systematic dependence on $\hbar$ \cite{moment}.  
However, the (much smaller) differences between the quantum and 
classical probabilities scale as $\hbar^2$, indicating that they are truly of 
quantal origin.

Our interest in Hyperion was stimulated by Zurek's provocative 
paper \cite{zurek2}.  He begins with an estimate of the break time, $T_E$, 
beyond which the Ehrenfest criterion of classicality will fail.  Since 
the width of a wave packet in a chaotic system grows exponentially 
with the Lyapunov exponent $\lambda$, the time taken for it to 
reach the scale $L$ over which the potential varies (typically of 
order of the system size) will be $T_E = \lambda^{-1} \ln(L/\Delta 
x_0)$, where $\Delta x_0$ is the initial width of the wave packet.  
Zurek chooses $\Delta x_0 = \Delta p_0/\hbar$, with the width of 
the momentum distribution $\Delta p_0$ being estimated from thermal 
fluctuations, and thereby obtains the now-familiar result that the 
break time $T_E$ scales as $\ln(\hbar^{-1})$.  This $T_E$ can be 
quite short, even for a macroscopic system, and because of the 
logarithmic dependence, it is not sensitive to detailed assumptions 
about the initial state.  Thus, according to the Ehrenfest criterion 
for 
classicality, the tumbling motion of Hyperion should long ago have 
ceased to be classical.

This paradoxical conclusion is not affected by including 
environmental decoherence.  
Decoherence converts a pure state into a mixed state, but it 
does not produce localization of the position probability density, and 
so it has no significant influence on the breakdown of Ehrenfest's 
theorem.  (Indeed, the diffusive term that describes the effect of the 
environment in the master equation for the density matrix will have a slight 
delocalizing effect.)  Therefore, if the Ehrenfest criterion were the 
sole criterion for classicality, we would still be faced with the 
paradoxical conclusion that macroscopic bodies like Hyperion should 
be grossly non-classical.

In \cite{zurek,zurek2}  Zurek next considers the 
equation of motion of the Wigner function, which has the form of the 
classical Liouville equation plus a series of $\hbar$-dependent terms, 
called the \textit{Moyal terms}.  The limit of the Liouville regime of 
classicality will presumably be reached when the Moyal terms have 
a significant effect.  Unfortunately, Zurek does not distinguish 
between the Ehrenfest and Liouville regimes, and denotes both break 
times as $t_{\hbar}$.  This is a serious confusion, since they are both 
conceptually and numerically distinct.

While the time limit $T_E$ of the Ehrenfest regime is easy to 
estimate reliably, an analogous limit for the Liouville regime is much 
more difficult to obtain.  Zurek uses heuristic arguments to claim 
that decoherence tends to counter the effects of the Moyal terms, and 
he obtains a Liouville break time similar in form to $T_E$ \cite{zurek}.  
We consider this conclusion to be doubtful for several reasons.  
Habib et al. \cite{habib} have shown that, 
in order for the non-negativity of the density matrix (``rho-positivity" 
in their terminology) to be preserved, the Moyal terms must have 
subtle effects that are not counteracted by decoherence.  Full 
numerical computations for a driven system \cite{intspin} and for an autonomous 
system \cite{2rotors} have found 
that, although the differences between quantum and classical 
averages of observables have a brief period of exponential growth, 
these differences reach a saturation value, about which they 
fluctuate irregularly.  This saturation value is much smaller than the 
system size (unlike the deviations from Ehrenfest's theorem), and it 
tends to zero as some small power of $\hbar$.  The break time 
estimated by Zurek would be relevant only if the exponential growth 
of these QC differences continued until they reached macroscopic 
size.

In this paper we perform classical and quantum computations for the 
chaotic tumbling of an object like Hyperion, which complement the 
published classical theory \cite{wisdom}.  The actual value of the de 
Broglie wavelength is, of course, much too short to be treated in a 
numerical integration of the Schr\"{o}dinger equation.  So we cast the 
equation into dimensionless form, and solve it for a range of the 
dimensionless $\hbar$ parameter.  These results lead to scaling 
relations, from which we can extrapolate to estimate the values 
appropriate to Hyperion.  We study the system with and without 
environmental decoherence, discuss the conditions 
needed to ensure classicality, as well as examine whether there is a qualtiative difference between the classical limit of regular and chaotic initial states.

\section{Model}

\begin{figure}[tp]
\centering
\scalebox{.4}{\includegraphics{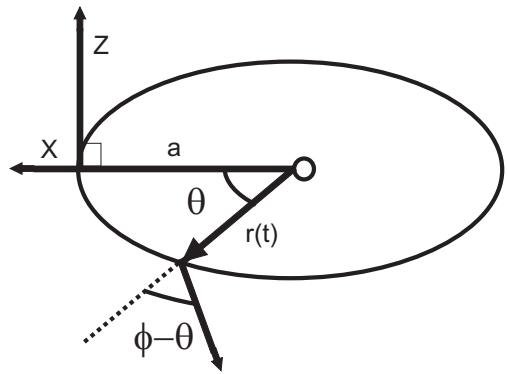}}
\caption{\label{orbit}Orbit of satellite spinning about the z-axis 
perpendicular to the orbital plane.  $\theta$ denotes the position of the 
satellite on the orbit, and $\phi$ is the orentation of the satellite 
with respect to the semi-major axis of the orbit.}
\end{figure}

Our model of Hyperion's rotation was first suggested in 1988 by Wisdom
 \cite{wisdom}.  It assumes that Hyperion's center of mass travels in 
an elliptical orbit about Saturn, and that its orbit is unaffected by its 
rotation.  However, since Hyperion is an extended object, Saturn's 
gravitational field is not constant over its volume.  Since its  mass 
distribution is not spherical, the variation in the gravitational field can 
produce a net torque.  To lowest order in a multipole expansion of the mass 
distribution, this torque depends on the quadrupole moment of the mass 
distribution, and for simplicity we neglect all higher order moments.

 It should be noted that this configuration is attitude unstable,
 and so small inclinations of $I_3$ towards the plane of the orbit will 
 tend to grow.
  However this simplifying assumption makes both the classical and the
 quantum mechanical computations feasible.

The coordinate system is shown in Fig. \ref{orbit}. 
 The space-fixed x-axis is along the semi-major axis of
 the orbital ellipse, and the z-axis is perpendicular to the orbital plane. 
 The angle $\theta$ denotes the position of the satellite in the orbit. 
 The axis of the smallest moment of inertia ($I_1$), makes an angle $\phi$ 
with respect to the x-axis, hence the angle between the body axis
 of $I_1$ and the radius vector $\vec{r}$ is $\phi-\theta$. 
 The largest moment of inertia $I_3$ is parallel to the z-axis.
 The canonical coordinates for this system are the angular momentum
 and the orientation of the satellite $\{L_z,\phi\}$. 

The coupling of the gravitational field to the satellite is obtained
by a Taylor expansion of the potential about the satellite's center of mass,

\begin{equation}
\label{Ham1}
H=\frac{L_z^2}{2I_{zz}}+\left.\sum_i\sum_j \int \rho(\vec{r})x_i x_j d^3x \frac{\partial^2V}{\partial x_i\partial x_j}\right|_{\vec{x}=0}+\cdots
\end{equation}
Here $x_i$ refers to the distance along the $i$'th space-fixed axis 
from the center of mass of the satellite.  

\begin{equation}
\left. \frac{\partial^2V}{\partial x_i \partial x_j}\right|_{\vec{x}=0}=diag \left[\frac{2Gm}{r^3},\frac{-Gm}{r^3},\frac{-Gm}{r^3}\right]
\end{equation}
Here m is the mass of the gravitational source (Saturn), 
and r is the distance from the source to the satellite.
The first order term in \eqcite{Ham1} vanishes because the expansion is
about the center of mass, and the second order term 
is related to the moments of inertia tensor,

\begin{equation}
\label{moments}
I_{ij}=\int \rho [x^2\delta_{ij}-x_i x_j]d^3x
\end{equation}

Using Kepler's third law, which states $GM/a^3=4\pi^2/T^2$,` the Hamiltonian becomes

\begin{equation}
H=\frac{L_z^2}{2I_{3}}-\frac{3 \pi^2}{T^2}\left(\frac{a}{r(t)}\right)^3(I_{2}-I_{1})\cos(2[\phi-\theta(t)])
\label{hamexp}
\end{equation}
Here $T$ is the orbital period, $a$ is the length of the semi-major axis of the orbit, $L_z$ is the angular momentum about the z-axis, $I_3=I_{zz}$ is the moment of inertia for rotations in the orbital plane, and r(t) and $\theta(t)$ are the orbital coordinates of the satellite, which are functions of the period $T$ and the eccentricity $e$.  These functions are found by numerically integrating the equations of motion for the center of mass, using the code provided in \cite{Caroll}.


\subsection{Classical Equation Of Motion}

It is convenient to express the equation of motion in terms of 
dimensionless variables.  We introduce the anisotropy parameter,
\begin{equation}
\alpha=\frac{(I_{2}-I_{1})}{I_{3}}
\label{alpha}
\end{equation}
a dimensionless time (in units of the orbital period),
\begin{equation}
\tau=t/T
\label{tau}
\end{equation}
and a dimensionless angular momentum $J_z$ in terms of the dimensional angular momentum $L_z$
\begin{equation}
J_z=\frac{L_zT}{I_3}
\end{equation}

To estimate $\alpha$ for Hyperion, we use the observed lengths of 
its principle axes (410 $\pm$ 10, 260 $\pm$ 10, 220 $\pm$ 10 km)
 \cite{Soderblom}, and assume that it is an ellipsoid of uniform mass density.  Hence

\begin{equation}
I_3=\frac{M}{5}(r_1^2+r_2^2)
\label{ellipmom}
\end{equation}
Here $r_i$ is half the length of the $i^{th}$ principle axis of the ellipsoid.
The other moments of inertia are obtained by cyclically permuting the indices. 
 Substituting \eqcite{ellipmom} into \eqcite{alpha} yields

\begin{equation}
\alpha=\frac{r_1^2-r_2^2}{r_1^2+r_2^2}
\end{equation}
Hence $\alpha = 0.43 \pm 0.04$.  In this work we used a slightly larger 
value, $\alpha=0.5$, because it leads to a more purely chaotic motion, 
whereas for $\alpha=0.43, e=0.1$ there are large regular islands embedded 
in the chaotic sea.  We wish to compare chaotic motions with regular motions,
and the differences would be obscured by a mixed phase space.

Following Wisdom \cite{wisdom}, we obtain the equation of motion 
(in dimensionless variables) to be
\begin{eqnarray}
\dot{\phi}&=&J_z\nonumber\\
\label{cleq2}
\dot{J_z}&=&-6 \pi^2\left(\frac{a}{r(\tau)}\right)^3\alpha\sin(2[\phi-\theta(\tau)])
\end{eqnarray}


\subsection{Quantum Mechanics}
The quantum mechanics will be solved by integrating the Schr\"{o}dinger
 equation in angular momentum representation. 
 The state vector is written as

\begin{equation}
\ket{\psi(t)}=\sum_m c_m(t)\ket{m} ,
\label{schrodinger}
\end{equation}
with $\ket{m}$ being an angular momentum eigenstate. 
 The matrix elements of the Hamiltonian are

\begin{eqnarray}
\bra{m}\hat{H}\ket{\psi}=\frac{\hbar^2m^2c_m}{2I_3}-\sum_{n}\frac{3\pi^2}{2T^2}c_n \left(\frac{a}{r(t)}\right)^3\times\nonumber\\(I_2-I_1)\frac{1}{2\pi}\int_0^{2\pi}(e^{i(n-m+2)\phi}e^{-2i\theta}+e^{i(n-m-2)\phi}e^{2i\theta})d\phi\nonumber\\
\label{qmeq2}
\end{eqnarray}
Using (\ref{qmeq2}) and (\ref{schrodinger}), the matrix equation
 $\bra{m}\hat{H}\ket{\psi}=i\hbar\bra{m}\frac{\partial}{\partial t}\ket{\psi}$
 becomes

\begin{eqnarray}
i\hbar\diff{c_m(t)}{t}=\frac{\hbar^2m^2c_m}{2I_3}-\frac{3\pi^2}{2T^2}\left(\frac{a}{r}\right)^3(I_2-I_1)\times\nonumber\\(c_{m+2}e^{2i\theta(t)}+c_{m-2}e^{-2i\theta(t)})
\label{qmeq3}
\end{eqnarray}

In addition to the dimensionless parameters $\tau$ and $\alpha$,
we now introduce a dimensionless  $\hbar$ parameter,

\begin{equation}
\beta=\frac{\hbar T}{I_3}
\label{beta}
\end{equation}
The dimensionless Schr\"{o}dinger equation then becomes

\begin{eqnarray}
i\diff{c_m}{\tau}=\frac{\beta m^2 c_m}{2}-\frac{3\pi^2}{2}\frac{\alpha}{\beta}\left({\frac{a}{r(\tau)}}\right)^3\times\nonumber\\(c_{m+2}e^{2i\theta(\tau)}+c_{m-2}e^{-2i\theta(\tau)})
\label{qmeq}
\end{eqnarray}

A peculiar feature of \eqcite{qmeq} is that the coefficient $c_m$ depends 
only on $c_{m+2}$ and $c_{m-2}$, therefore the even $c_m$ cannot interact 
with odd $c_m$.  This coupling arises from the invariance of the Hamiltonian
under rotations by $\pi$. But octapole and other odd moments are not invariant
 under rotations by $\pi$, so this symmetry is an artifact of the model. 


\subsection{Initial State}

The initial quantum state is chosen to be a Gaussian in angular momentum,

\begin{equation}
\label{initstate}
\ket{\psi}=\sum_{m}\exp\left(-\frac{(\beta m-J_0)^2}{2 \delta^2}-i\phi_0 m\right)\ket{m}
\end{equation}

Here $\beta m$ is a dimensionless angular momentum, $J_0$ is the 
average of the dimensionless angular momentum in the state, $\delta$ is 
its standard deviation, and $\phi_0$ is the 
central angle of the initial state. 
 These parameters will be varied to ensure that the initial states are 
in regions of phase space that are either purely chaotic or purely regular.

In principle the sum is from $m = -\infty$ to $+\infty$, but in practice 
it is restricted to a range $\{-K\cdots K\}$. 
The value of $K$ must be chosen so that this range includes all of the 
values of $J_z$ that have significant amplitudes in the time-dependent state. 
By examining phase-space diagrams for the classical distributions, 
we found that $|J_z|<20$ for all time, and so $K=20/\beta$ was sufficient
 to contain the quantum distribution.

The initial classical probability distributions are chosen so that they 
match the angular momentum and angular distributions for the initial 
quantum state.  
Because the initial state is a minimum uncertainty state with fixed 
width in angular momentum, its width in angle is proportional to $\beta$. 
Thus $\beta$ (dimensionless $\hbar$) enters into the classical calculation
 to ensure that the initial quantum and classical states correspond to
 each other.

\section{\label{regular}Results For A Non-Chaotic State}

\begin{figure}[htp]
\centering
\scalebox{.7}{\includegraphics{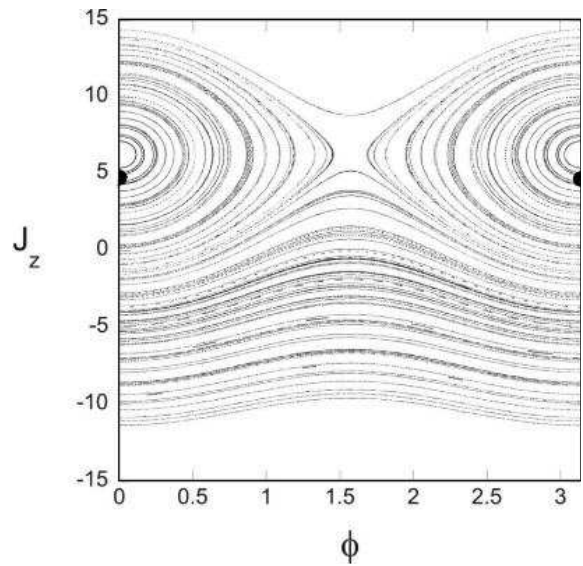}}
\caption{\label{rpoin}Poincare Section for $e=0$, $\alpha=0.5$.  Black circle denotes a typical initial state.}
\end{figure}
The classical limit of the quantum tumbling of a satellite will now be
examined for a non-chaotic state, to determine whether there is a
 qualitative difference between chaotic and non-chaotic systems in 
their approach to classicality.

Non-chaotic motion is ensured by choosing a circular orbit:
 $e=0$, $r(\tau)=a$, $\theta(\tau)=2\pi\tau$.  
The time dependence in \eqcite{cleq2} can be transformed away by the
 substitution $\Phi=\phi-2\pi\tau$, yielding an integrable equation of motion,

\begin{equation}
\label{cleq}
\ddot{\Phi}=-6 \pi^2\alpha\sin(2\Phi)
\end{equation}
Fixed points for this equation occur at the angles
 $\Phi=0,\frac{\pi}{2},\pi,\frac{3\pi}{2}$.
These fixed points describe motions in which Hyperion presents the same face 
to Saturn at all times.  The stable fixed points correspond to the
 smallest moment of inertia pointing towards Saturn.  

The initial state was chosen to be far from the unstable fixed point. 
 It is centered at $J_0=4$, with a standard deviation in $J_z$ of
 $\sigma=\frac{1}{\sqrt{2}}$ (see Fig. \ref{rpoin}), and a central anglei
 $\phi_0$ equal to zero.  

\begin{figure}[btp]
\centering
\scalebox{.7}{\includegraphics{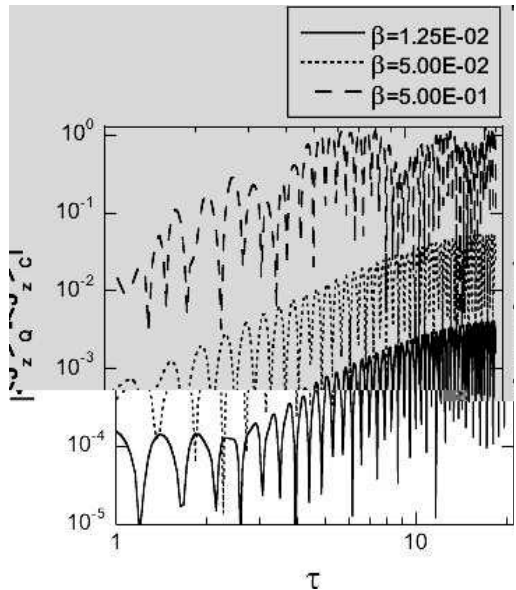}}
\caption{\label{regscale}QC differences in $\avg{J_z}$ vs $\tau$,
for several $\beta$, with $e=0$, $\alpha=0.5$. 
For $\beta=\{0.0125,0.05,0.5\}$ the statistical errors are 
 $\sigma_m =\{0.0002,0.0007,0.002\}$.}
\end{figure}

\begin{figure}[tbp]
\centering
\scalebox{0.6}{\includegraphics{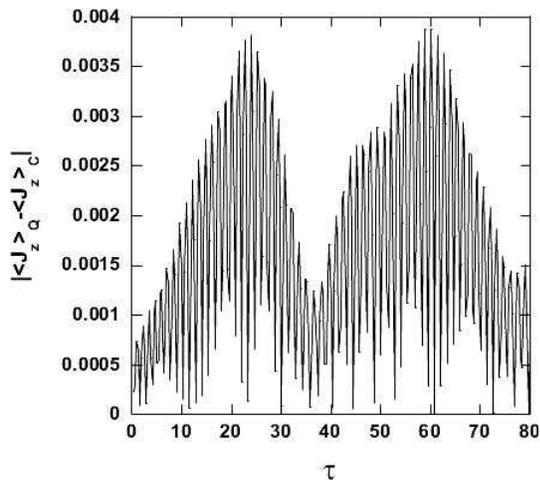}}
\caption{\label{bigscale}QC differences in $\avg{J_z}$ for $\beta = 0.0125$,
 with $e=0$, $\alpha=0.5$,  $\sigma_m=0.7\times 10^{-3}$}.
\end{figure}


\subsection{QC Differences in $\avg{J_z}$}
\begin{figure}[tbp]
\centering
\scalebox{0.7}{\includegraphics{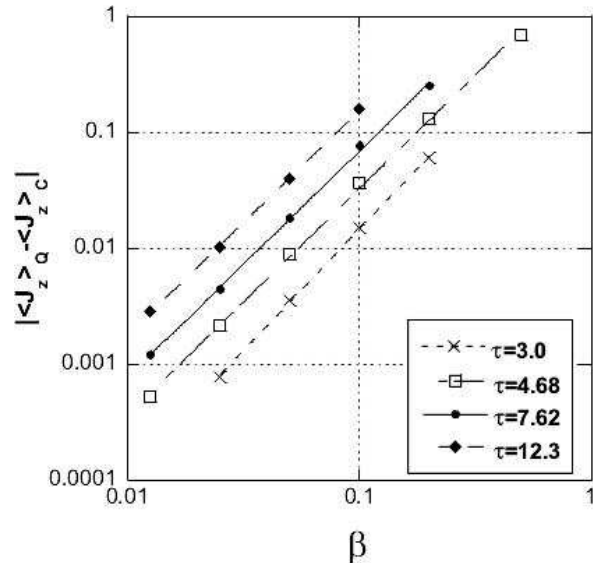}}
\caption{\label{betscale}Scaling of the early time QC differences 
in $\avg{J_z}$ with $\beta$, for $e=0$, $\alpha=0.5$. 
Showing a $\beta^2$ dependence.}
\end{figure}
The classical probability distributions are found by time evolving a finite
ensemble of systems, using \eqcite{cleq2}.  The distributions of $J_z$ and
 $\phi$ are found by randomly choosing the angular momentum and orientation 
of each member of the ensemble from probability distributions in
 $J_z$ and $\phi$ that correspond to the initial quantum state.

The finiteness of the ensemble leads to statistical errors, which may be
 reduced by increasing its size.  The standard deviation of the 
fluctuations in the mean is
\begin{equation}
\label{fluctuate}
\sigma_{m}=\frac{\sigma}{\sqrt{n}}
\end{equation}
Here $n$ is the number of members in the ensemble, $\sigma$ is the
 standard deviation of the distribution. 
Any difference between the computed mean values of the quantum and the 
classical variables is not significant unless it is larger than $\sigma_{m}$. 
 Ensembles of 1,000,000 to 20,000,000 particles were used to ensure that 
the typical QC differences are greater than $\sigma_m$.

As $\beta \rightarrow 0$ the QC differences become smaller, and thus 
a larger ensemble is needed to reduce the statistical errors below that level.
Hence different ensemble sizes were used for different values of $\beta$ 
in Fig. \ref{regscale}. 
The ensemble sizes were chosen so that $\sigma_m=\{0.0002,0.0007,0.002\}$ for $\beta=\{0.0125,0.05,0.5\}$.

Ensembles were evolved for several values of $\beta$, ranging from $\beta=0.5$ 
to $\beta=0.002$.  For $\beta<0.01$ the QC differences were far smaller than
 $\sigma_m$ for any computationally feasible ensemble sizes, so no data 
will be presented for $\beta<0.01$.  

A plot of the QC differences in $\avg{J_z}$ is shown in Fig. \ref{regscale}.
In this and similar figures, any QC differences smaller than $\sigma_m$ 
should be ignored, since they are dominated by statistical errors.
The QC differences oscillate on the scale of the driving force, and only
the envelope of these oscillations is of interest. 
From Fig. \ref{regscale} it is apparent that at early times the envelope of 
the QC differences grows as $\tau^{2}$.
For longer times the envelope of the QC differences is oscillatory,
as can be seen in Fig. \ref{bigscale}.  Such recurrences are typical 
for non-chaotic systems \cite{Haake}.
Fig. \ref{betscale} shows that, for fixed times, the QC differences in
 $\avg{J_z}$ scale as $\beta^2$.  This result is similar to that found for 
some other systems \cite{moment}.

\begin{figure}[tb]
\centering
\scalebox{0.7}{\includegraphics{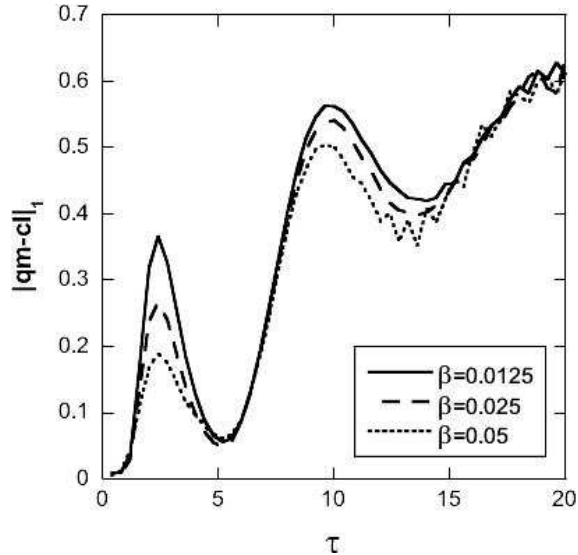}}
\caption{\label{regabs}Variation of \abdiff [\eqcite{abdiff}] with time and $\beta$ for the nonchaotic state ($e=0$, $\alpha=0.5$).  Each classical ensemble has 1,000,000 members.}
\end{figure}


\subsection{QC Differences in Distributions}

The differences in $\avg{J_z}$ alone are insufficient to fully describe the 
differences between quantum and classical systems because two different 
probability distributions can have the same mean but different variances 
and higher moments.  We shall now examine the differences between probability 
distributions, and how they scale with $\beta$.

Since the angular momentum distributions are discrete, one can regard them 
as vectors, and measure the difference between the quantum and classical 
probability vectors by the 1-norm, defined as

\begin{equation}
\label{abdiff}
\abdiffm=\sum_{m} |P_{cl}(m)-P_{qm}(m)|
\end{equation}
This probability distribution is normalized so that  $\sum_m P(m)=1$. 
Alternatively, one can define a probability density, which is normalized so that
 $\int\tilde{P}(\hbar m) d(\hbar m)=1$. 
 Then the 1-norm of the probability densities takes the form 
\begin{equation}
\abdiffm=\int_{-\infty}^{\infty}|\tilde{P}_{cl}(\hbar m)-\tilde{P}_{qm}(\hbar m)|d(\hbar m)
\end{equation}
These two forms are equivalent because $\tilde{P}(\hbar m)= P(\hbar m)/\hbar$,
and the additional factor of $\hbar$ is cancelled by the factor of $\hbar$ 
 in the integral.



Fig. \ref{regabs} shows that the QC differences in the probability distributions
do not tend to zero as $\beta \rightarrow 0$.  
This lack of pointwise convergence of the quantum probability distributions 
to the classical limit has also been observed for other systems,
such as a particle in a box and the kicked rotor \cite{fractal}. 
In these one-dimensional driven system, the quantum probability distributions 
develop a fractal-like structure, and only the smooth background converges
to the classical probability distribution.
We will show in section \ref{smoothing} that a similar result holds 
for Hyperion.

\section{\label{chaotic}Results For a Chaotic State}

\begin{figure}[htp]
\centering
\scalebox{.76}{\includegraphics{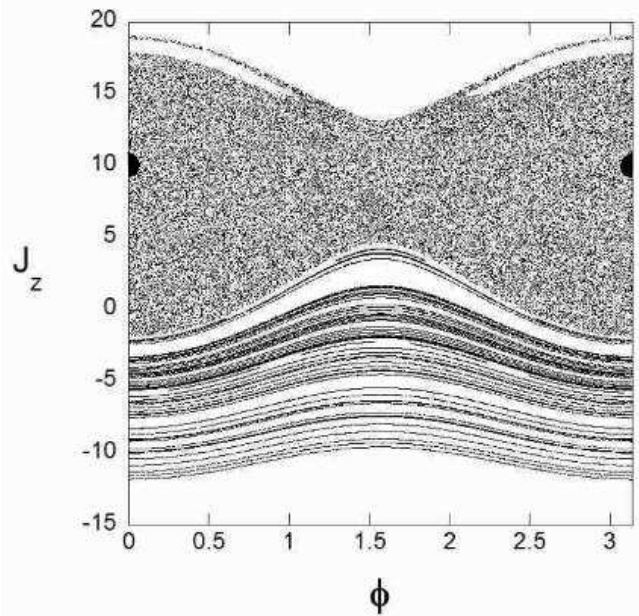}}
\caption{\label{chpoin}Poincare Section for a chaotic state, $\alpha=0.5$, 
$e=0.1$.  Black circle denotes a typical initial state, with $J_z=10$ and
$\delta=0.5$}
\end{figure}

In this section the rotation of a satellite is investigated for a chaotic
 state.  The previous value of $\alpha=0.5$ is used, but now the
 eccentricity is taken to be Hyperion's value of $e=0.1$.  
The computation is carried out as in section \ref{regular}.

The initial state is centered at dimensionless angular momentum $J_0=10$,
with a standard deviation of $\sigma=0.5$, and a central angle $\phi_0=0$.  
This state is in the chaotic sea, far away from any regular torii, as can be 
seen in Figure \ref{chpoin}.  The maximum Lyapunov exponent for the chaotic sea is $\lambda=0.85$.


\begin{figure}[tbp]
\centering
\scalebox{0.7}{\includegraphics{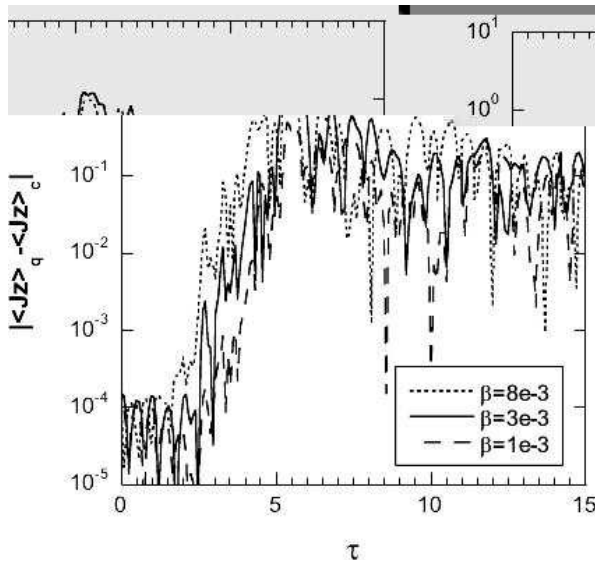}}
\caption{\label{cmeans}QC differences in $\avg{J_z}$ vs $\tau$ for a 
chaotic state, $\alpha=0.5, e=0.1$. 
For $\beta=\{1\times 10^{-3},3\times 10^{-3},8\times 10^{-3}\}$ the 
statistical errors are 
$\sigma_m= \{1.4\times 10^{-4},1.4\times 10^{-4},2.2\times 10^{-4}\}$.}
\end{figure}

\begin{figure}[tbp]
\scalebox{0.8}{\includegraphics{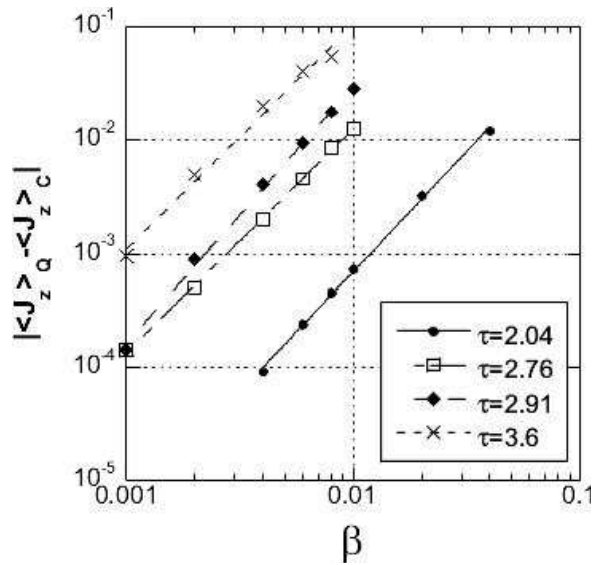}}
\caption{\label{csides}QC differences in $\avg{J_z}$ vs $\beta$, 
for a chaotic state before saturation is reached,
showing a $\beta^2$ dependence.}
\end{figure}

This state was evolved for several periods of the driving force, and 
the differences between the quantum and classical results were computed. 
 In Fig. \ref{cmeans} the QC differences in $\avg{J_z}$ are initially 
dominated by statistical errors, which are approximately
 $\sigma_m=2\times 10^{-4}$.  The QC differences in $\avg{J_z}$ grow 
exponentially with time until they saturate at $\tau \approx 6$. 
 This saturation occurs when the classical trajectories ergodically fill
 the chaotic sea.  
For $\tau>20$ the classical ensemble saturates at $\avg{J_z}\approx 8.2$. 
The quantum value of $\avg{J_z}$ also saturates at approximately the same value,
 but with irregular fluctuations superimposed. 
This suggests that the QC differences here are dominated by quantum 
fluctuations, once the probability distributions have saturated the chaotic sea.


\subsection{QC Differences in $\avg{J_z}$ for Early Times}

To determine how the QC differences scale with $\beta$, we varied $\beta$ 
with $\tau$ fixed at the times of the peaks in Fig. \ref{cmeans}. 
 From Fig. \ref{csides} it can be seen that the QC differences in $\avg{J_z}$ 
scale as $\beta^2$; the same scaling as was found for the non-chaotic states.
This $\beta^2$ scaling was previously found by a different method for 
other systems \cite{moment}. 
The classical ensemble sizes were chosen so that 
 $\sigma_m=\{1.6\times 10^{-4},1.6\times 10^{-4},2.2\times 10^{-4}\}$ 
for $\beta=\{1\times 10^{-3},3\times 10^{-3},8\times 10^{-3}\}$.

In the initial growth region of Fig. \ref{cmeans},the QC differences 
in $\avg{J_z}$ vary with time as 
\begin{equation}
|\avg{J_z}_Q-\avg{J_z}_C|\propto e^{2.9\tau}
\label{expgrow}  
\end{equation}
for $\tau=2$ between 2 and 5.5.
The exponent in \eqcite{expgrow} appears to be independent of the value of
 $\beta$.  The exponent is greater than $2\lambda$, implying that the 
QC differences grow at a rate that is greater than the classical 
Lyapunov exponent.  Similar results have been obtained for some other 
systems \cite{moment,qmcldiff,intspin}. 

The exponential growth eventually saturates.  
This cessation of exponential growth of the QC differences in $\avg{J_z}$ 
is relevant to Zurek's argument \cite{zurek2} that, absent decoherence, the 
QC differences for Hyperion should reach macroscopic size within about 
20 years.  That argument implicitly assumes that the QC differences 
will continue to grow exponentially until they reach the size of the system.
However, we find that not to be the case.


\subsection{QC Differences in $\avg{J_z}$ for the Saturation Regime}

The maximum QC differences occurs in the saturation regime.
If these differences converge to 0 as $\beta \rightarrow 0$ 
then the classical limit will be reached for all times, and there will be
no \textit{break time} beyond which QC correspondence fails.

\begin{figure}[tbp]
\scalebox{0.7}{\includegraphics{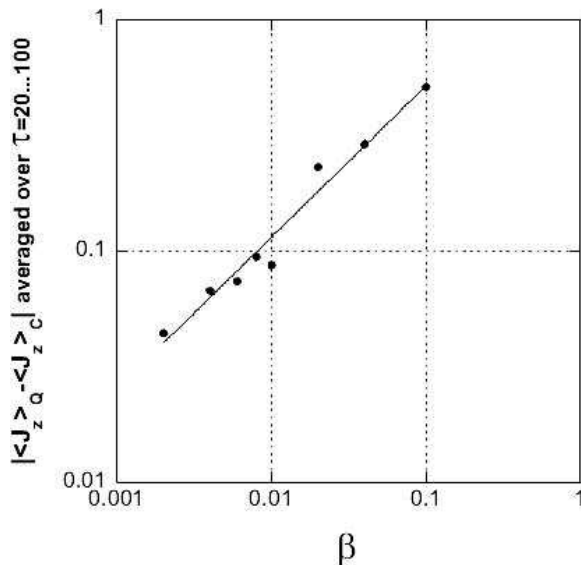}}
\caption{\label{discrep}QC differences in $\avg{J_z}$, averaged over 
$\tau$ from $20$ to $100$.  Initial state is in the chaotic sea. 
$e=0.1$, $\alpha=0.5$.  
Suggests that $\avg{|\avg{J_z}_Q-\avg{J_z}_C|}\propto\beta^{2/3}$ 
in the saturation regime.}
\end{figure}

At the beginning of the saturation region
 (Fig. \ref{cmeans} and Fig. \ref{nexpec}), the QC differences in
 $\avg{J_z}$ reach a maximum, before decaying to a 
saturation level, about which they fluctuate irregularly.  
Because of this fluctuation in the saturation regime, we calculate the 
time average of the QC differences.  Here $|\avg{J_z}_Q-\avg{J_z}_C|$ 
was averaged over $\tau$ from 20 to 100.  
As shown in Figure \ref{discrep}, these averaged QC differences tend to scale 
as $\beta^{2/3}$.

\begin{figure}[tbp]
\scalebox{0.7}{\includegraphics{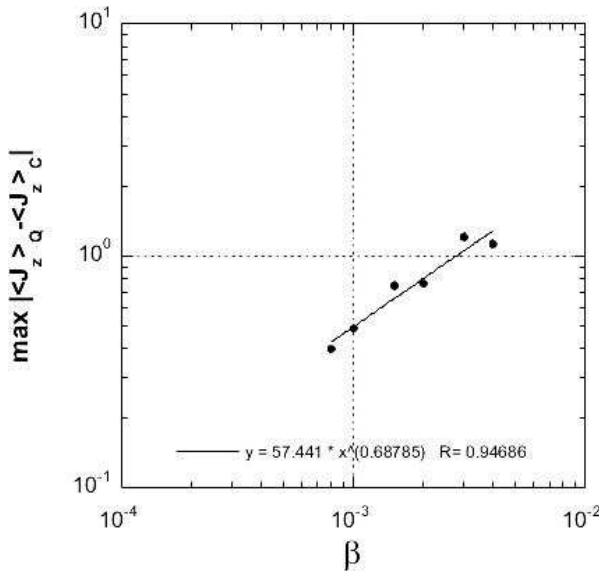}}
\caption{\label{maxscale}Maximum QC differences in $\avg{J_z}$ versus $\beta$ 
for a chaotic state. $e=0.1$, $\alpha=0.5$. 
Suggests that this maximun difference scales as $\beta^{2/3}$. }
\end{figure}

For sufficiently small $\beta$, the peak QC differences also scales 
as $\beta^{2/3}$ (see Fig. \ref{maxscale}). 
This scaling also was found for the maximum QC differences in a model of
coupled pendulums \cite{2rotors}, so it might be generic for chaotic systems 
in the saturation regime.

\subsection{QC Differences in Probability Distributions}

\begin{figure}[tbp]
\scalebox{.73}{\includegraphics{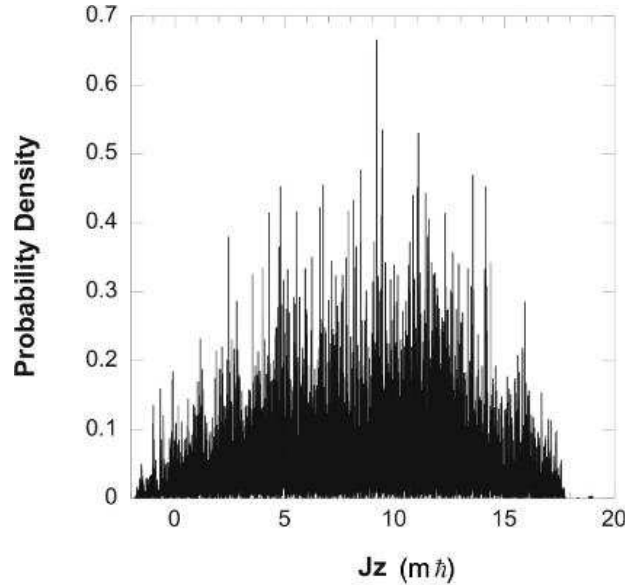}}
\caption{\label{qprob}Quantum probability density for $\tau=40.0$, $\beta = 0.002$}
\end{figure}

The probability distributions contain much more information than do
the averages of observables.  
These probability distributions are shown in Figures \ref{qprob} and \ref{501}.
We use the quantity \abdiff (defined in
 \eqcite{abdiff}) as a measure of the QC differences in the probabilities.

\begin{figure}[htbp]
\scalebox{0.71}{\includegraphics{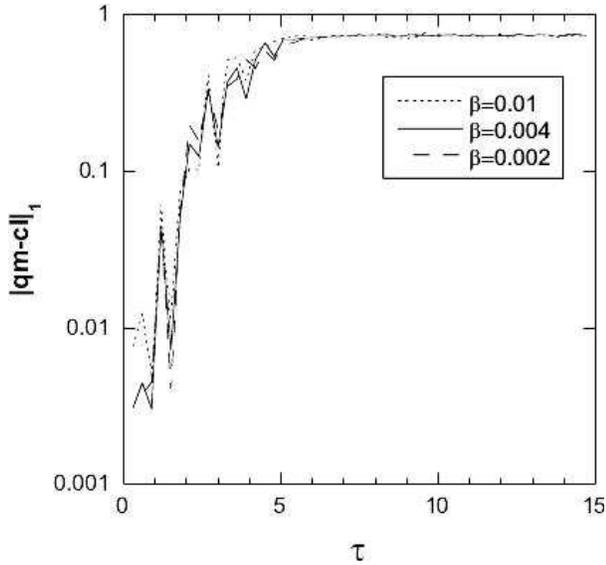}}
\caption{\label{c_abs}\abdiff vs time for different values of $\beta$, 
for a chaotic state, $e=0.1$}
\end{figure}

As can be seen from Fig. \ref{c_abs}, \abdiff increases with time 
before saturating, but fails to converge to 0 as $\beta \rightarrow 0$. 
 Since pointwise convergence does not occur, neither for the chaotic 
nor for the non-chaotic states, this lack of pointwise convergence is not
 a result of chaos.  

\begin{figure}[tbp]
\scalebox{.79}{\includegraphics{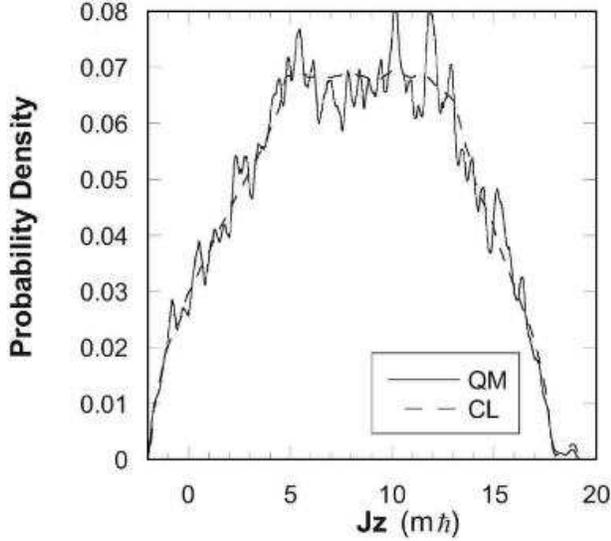}}
\caption{\label{coarse}Quantum and classical probability densities for $\tau=40.0$, 
$\beta = 0.002$.  Both quantum and classical densities are convolved with a triangular 
filter of width $0.25$ in $J_z$}
\end{figure}

Most of the QC differences in the probabilities occur on a very fine scale, 
and a modest amount of smoothing is sufficient to cause the quantum probability
distributions to better approximate the classical results.  
Fig. \ref{coarse} shows that the differences between the two distributions 
are dramatically reduced by smoothing them over a small width. 
This smoothing process is discussed and compared to environmental decoherence 
in section \ref{smoothing}.


To summarize the results of this section, for early times the QC differences 
in $\avg{J_z}$ scale as $\beta^2$, and increase exponentially with time.  
The exponential growth ceases when the probability distributions saturate the chaotic sea.
Both the maximum values and the saturation levels of the QC differences
were found to scale as $\beta^{2/3}$. 
A small amount of smoothing can dramatically reduce the QC differences in 
the probability distributions, since most of the differences come from
very fine scale structures in the quantum probability distributions.

\section{\label{decoh}Environmental Effects}

Interaction with the environment leads to decoherence and dissipation.  
Decoherence is a quantum effect that causes interference patterns to decay.  
The time scale upon which this happens is model dependant, and for some systems
there is no single decoherence timescale \cite{decondeco}. 
Strunz et al. \cite{univdeco} suggest that for the rapid decoherence expected 
in macroscopic bodies, the decay time varies as a small power of $\hbar$,
and is not sensitive to the system Hamiltonian.

\begin{figure}[tb]
\scalebox{.7}{\includegraphics{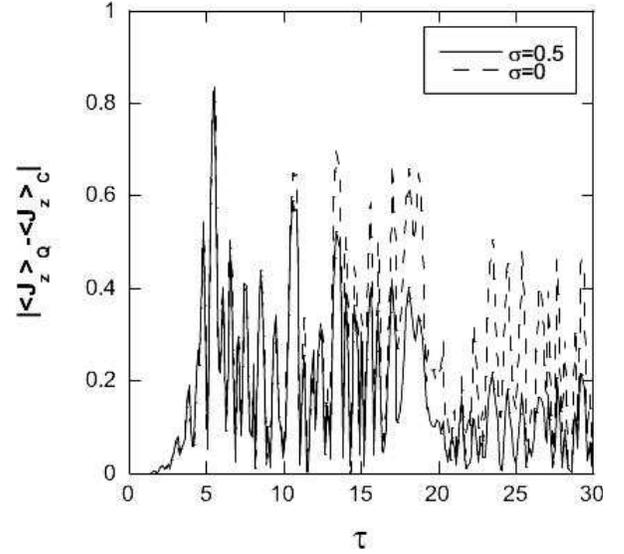}}
\caption{\label{nexpec}Variation of $\avg{J_z}$ with $\tau$ with and without the 
random potential with $\sigma/V_{ch}=0.024$, $\tau_c=0.01$. 
(chaotic state, $\beta=0.05$)}
\end{figure}

\begin{figure}[tb]
\scalebox{.7}{\includegraphics{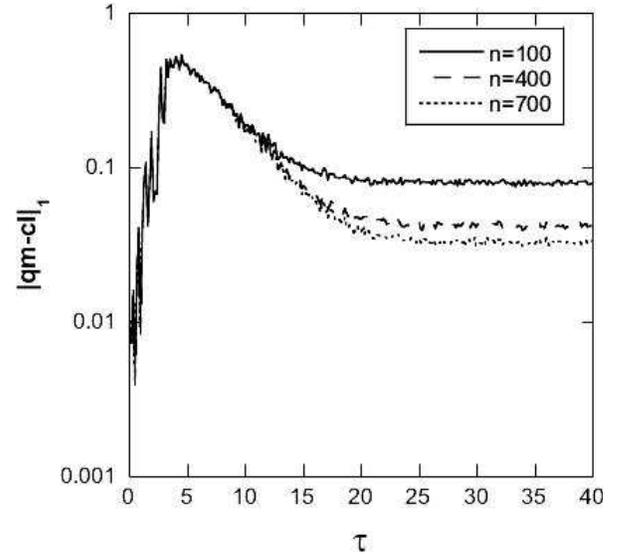}}
\caption{\label{rescale}Variation of \abdiff with $\tau$ for different numbers $n$ of 
realizations of the random potential with $\sigma/V_{ch}=0.012$, $\tau_c=0.01$.
(chaotic state)}
\end{figure}

\begin{figure}[tb]
\scalebox{.7}{\includegraphics{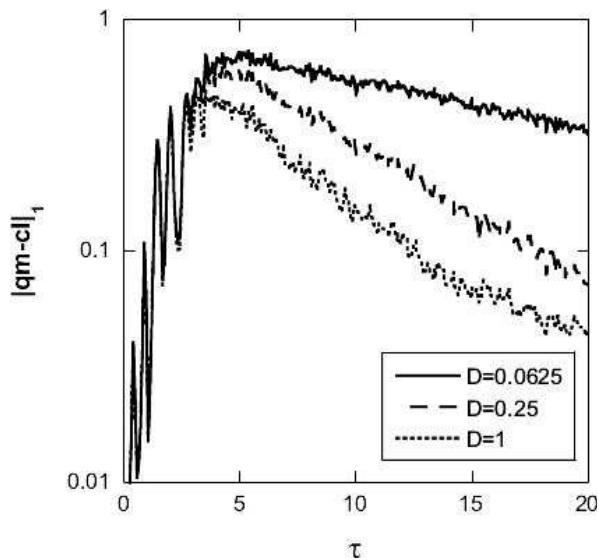}}
\caption{\label{sscale}\abdiff vs $\tau$ for $\beta=0.05$, for varying $D=\sigma^2\tau_c/6$.}
\end{figure}

\begin{figure}[tpb]
\scalebox{.7}{\includegraphics{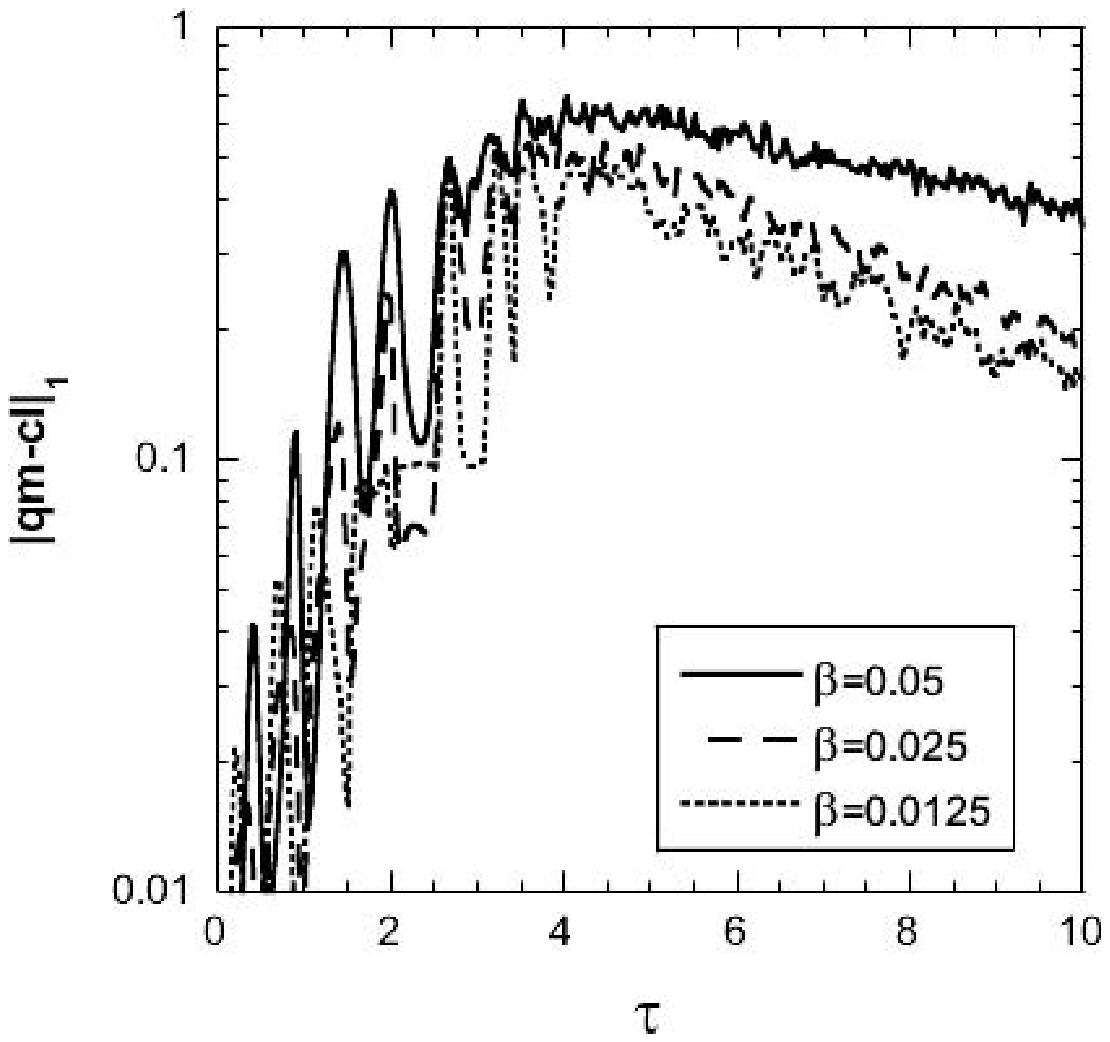}}
\caption{\label{bscale}\abdiff vs $\tau$ for varying $\beta$, with $\tau_c=0.01$ and 
 $\sigma/V_{ch}=0.012$, for a chaotic state.}
\end{figure}

\begin{figure*}[t!]
\scalebox{.7}{\includegraphics{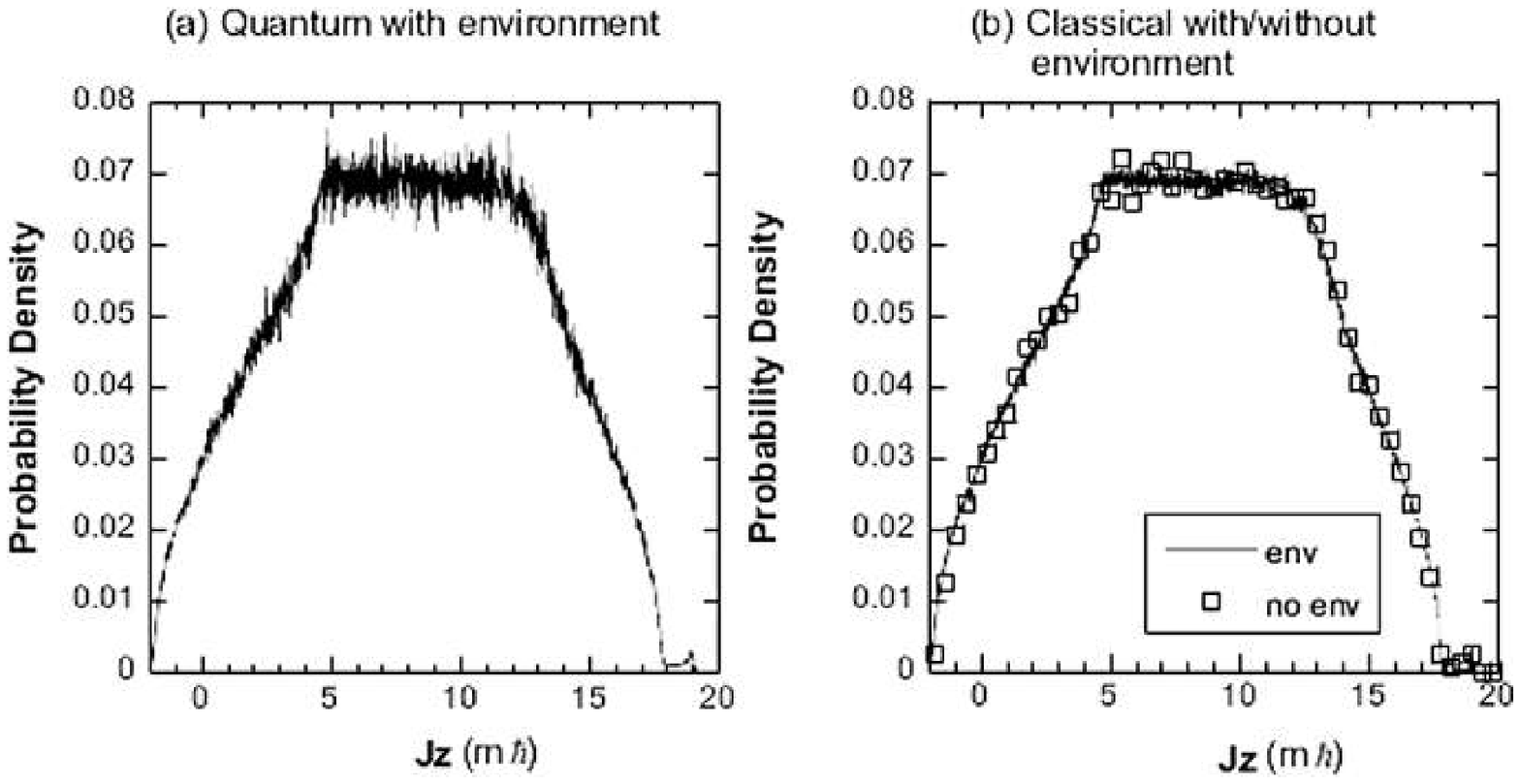}}
\caption{\label{501}Quantum and Classical probability distributions at $\tau=40$,
 with $\beta=0.0125$, $\sigma/V_{ch}=0.012$, and $\tau_c=0.01$ for the chaotic state.
In (b) the solid lines denote the results with the environment, and squares without 
the environment, showing that the classical probability distribution is not significantly 
affected by the environment.}
\end{figure*}

Dissipation is a classical effect which results in diffusive spreading
of the probability distributions.  Often, the timescale for dissipation 
is much longer than the timescale for decoherence, 
and dissipation is insensitive to $\hbar$, unlike decoherence.
But both effects are present together, and it is not always easy 
to separate them.

The effect of the environment on a quantum system is often treated by a 
master equation that has non-unitary time evolution.   
Because an initially pure state can evolve into a mixed state, it is necessary
to compute the density matrix, which requires much greater storage than does 
the computation of a state vector. The requirements for storage and 
computation time scale like $K^2$, where K is the number of basis vectors 
needed to store a state vector.

An alternative method is to perform $n$ evolutions of \eqcite{qmeq} 
with a different realization of the random potential added for each run.  
Averaging the probability distributions that result from the each of the $n$
runs is physically equivalent to tracing over the environmental variables.
The advantage of this method is that the computational resources for each 
run scale as $K$, rather than $K^2$ for the master equation.  
On the other hand, to achieve good accuracy, a large number of realizations 
of the random potential must be considered, in order to reduce the 
statistical errors in the quantum calculation.  However the number of 
realizations of the random potential that was needed to get sufficient
accuracy was considerably less than $K$, so this method was much more 
computationally efficient than integrating the master equation.

A stochastic potential is included, in both the quantum and classical mechanics,
 to model the effect of the environment on the satellite.  
The simplest stochastic potential that yields a random torque is

\begin{equation}
\label{ranpot}
H_{int}=V_0 R(t) \cos(\phi)
\end{equation}
Here $R(t)$ is a correlated random function of zero mean and unit variance,
 $V_0$ is the amplitude of the random potential,
 and $\tau_c$ is its correlation time. 
A correlated random function is used because the fluctuations in the
 environment do not occur instantly, but rather they occur and decay on
some time scale $\tau_c$.  
The correlated random sequence $R(t)$ can be constructed from an uncorrelated 
sequence, as is shown in Appendix A.
The results are not sensitive to the exact form of \eqcite{ranpot}, and
qualitatively similar results were obtained when  $\cos(\phi)$ was
 replaced by $\cos(2\phi)$.  
As is shown in Appendices B and C, the effects of the environment are 
expected to depend mainly on the product $\sigma^2 \tau_c$, rather than 
on the two parameters separately.  Therefore we label the results by the 
momentum diffusion parameter, $D = \sigma^2 \tau_c /6$, which is derived 
in Appendix B.

The environmental perturbation should be much weaker 
than the tidal force on the satellite. Hence we compare the interaction 
potential \eqcite{ranpot} to the amplitude of the tidal potential, made dimensionless by dividing by $I_3/T^2$,  
which is $V_{ch}=3\sqrt{2}\pi^2\alpha$,
 or $V_{ch}\approx 21$ for $\alpha=0.5$.  In all cases reported in this paper,
the environmental perturbation was so weak as to have no significant effect 
on the classical results, so its only significant effect is to produce 
decoherence in the quantum results.
The same parameters as in the previous chaotic case were used,
 $\alpha=0.5$ and $e=0.1$. 

Many realizations of the random potential were computed, and the results 
averaged, to get an accurate measure of the effects of the environment.
We used 500 realizations to obtain results that are not strongly affected
 by statistical errors.
Fig. \ref{nexpec} shows the QC differences in $\avg{J_z}$, with and without 
the random environmental potential.  The environment has no significant 
effect at early times, but in the saturation regime the QC differences are 
reduced. Since the primary effect of environmental decoherence is
 to destroy fine-scale structures in the probability distributions, 
which do not affect averages like $\avg{J_z}$, this result may seem surprising. 
In fact, a typical trace of $\avg{J_z}$ versus time for a single realization 
of the random potential will look very much like that from a run without 
the random potential in Fig. \ref{nexpec}.  But as time progresses, 
the oscillations in $\avg{J_z}$ for different realizations of the 
random potential tend to get out of phase with each other, and the decreased 
amplitudes of the QC differences in Fig. \ref{nexpec} are due to the 
averaging over the many different realizations of the random potential.

\begin{figure}[h!tpb]
\scalebox{.7}{\includegraphics{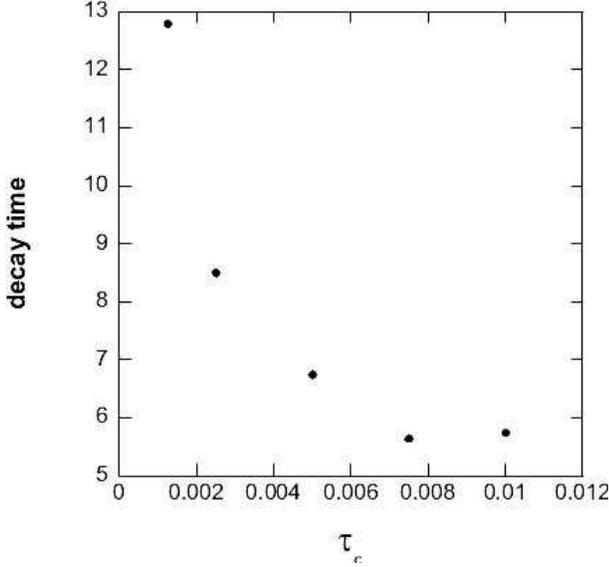}}
\caption{\label{decayt}Decay times $\tau_d$ of \abdiff vs the correlation time
 $\tau_c$ of the perturbing environment, with $\beta=0.05$, 
$\sigma/V_{ch}=0.012$, in the saturation regime ($\alpha=0.5$, $e=0.1$)}
\end{figure}

In Fig. \ref{rescale}, 100 realizations of the interaction potential were 
 sufficient to find the maximum QC differences in \abdiff (\eqcite{abdiff}).  
However, 700 realizations of the interaction potential were insufficient 
to resolve the QC differences in \abdiff  in the saturation 
regime, and so for computational reasons, these will be estimated rather than 
directly computed.  The variation of these differences with $\beta$ and 
the environmental parameters can be seen in Figures. \ref{sscale} and \ref{bscale}.

It can be seen in Fig. \ref{501} that, with the inclusion of the environmental 
perturbation, the quantum probability distribution is much closer to the 
classical distribution than without the environment (compare Fig. \ref{qprob}). 


\subsection{Environmental Effects in the Saturation Regime}

In the saturation regime, the QC differences in \abdiff (\eqcite{abdiff}) 
appear to decay exponentially from their maximum value to a saturation level
(see Fig. \ref{sscale} and \ref{bscale}).  
The rate of this decay is a duffusion time, and is not the decoherence time, 
as will be shown shortly.  From Fig. \ref{decayt} it is clear that, for 
sufficiently large values of $\tau_c$, the decay time $\tau_d$ no longer 
depends on $\tau_c$, but settles at $\tau_d \approx 5.6$.  
For sufficiently large $\sigma$ and sufficiently small $\beta$, the decay time 
was found to also have approximately this same limit.

\begin{figure}[tbp]
\scalebox{.7}{\includegraphics{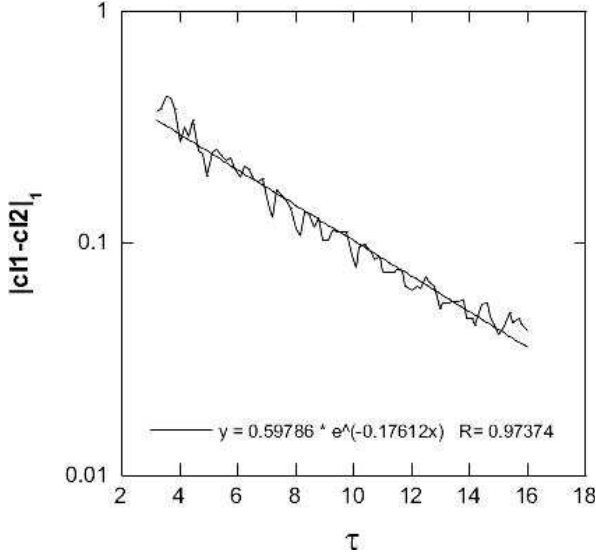}}
\caption{\label{decayt2} 1-norm of the difference between two classical 
ensembles, one with $\avg{J_z}=10$, the other $\avg{J_z}=11$. 
 $\beta=0.05$, $\sigma/V_{ch}=0.012$, $\tau_c=0.01$}
\end{figure}

To test whether this decay rate is governed by quantum mechanics, we compared 
two classical ensembles with different initial values of $J_0$
($J_0=10$ and $J_0=11$), and computed the 1-norm of the difference between 
them as a function of time.  Fig. \ref{decayt2} shows that these initially 
different classical ensembles converge at a rate given by $\tau_d=5.6$.  
So, apparently, this time scale measures how quickly the differences between 
two different distributions decrease as they both grow to fill the chaotic sea.

\begin{figure}[htb]
\scalebox{.7}{\includegraphics{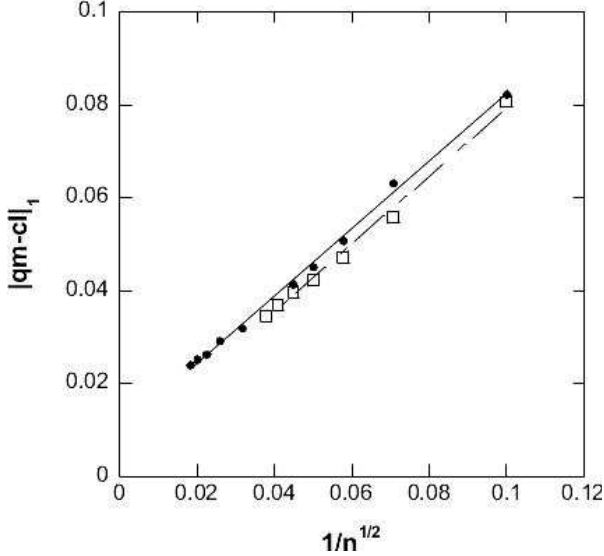}}
\caption{\label{const} \abdiff vs $n^{-1/2}$ at $\tau=40$. $n$ is the number 
of realizations of the random potential. 
($\sigma/V_{ch}=0.012$, $\tau_c=0.01$ for the chaotic state)}
\end{figure}

It is not clear from Fig. \ref{sscale} and \ref{bscale} whether the QC differences 
in the probability distributions eventually decrease to zero or reach 
a non-zero long-time limit.
In Fig. \ref{const} the long-time saturation level of \abdiff is plotted 
as a function of the number $n$ of realizations of the random potential.
In the limit $n \rightarrow \infty$ the QC differences approach a small value 
that appears to be slightly positive.  However that extrapolated limit is 
substantially smaller than the typical statistical errors for ensemble sizes 
of 1,000,000 to 2,000,000, and so is not significantly different from zero.  

\begin{figure}[tbp]
\scalebox{.7}{\includegraphics{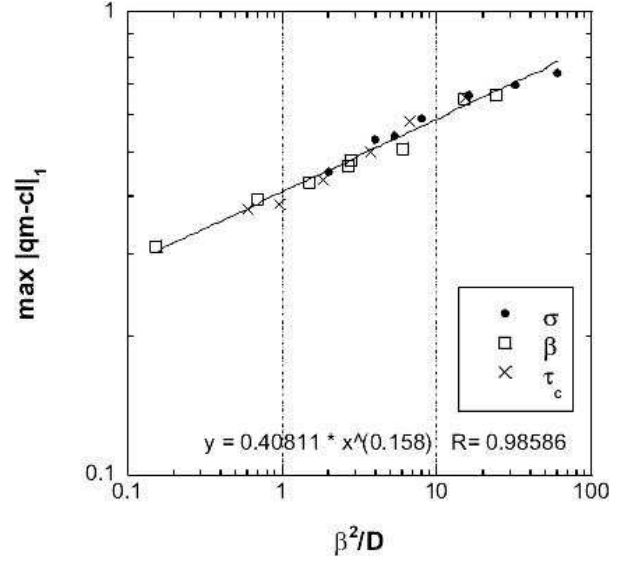}}
\caption{\label{paramscale}Maximum values of \abdiff vs $\beta^2/D$.  
The points labeled $\beta, \sigma, and \tau_c$ represent data sets where $\beta, \sigma,$ 
and $\tau_c$ were varied with the other two parameters held constant.}
\end{figure}


\subsection{Scaling of the Maximum QC Differences}

The maximum value of \abdiff for the quantum and classical probability 
distributions must depend on the three parameters $\beta$, $\sigma$, and 
$\tau_c$.  
However, in agreement with arguments presented by Pattanayak et al. 
\cite{Patyak}, the data was found to collapse 
onto a single curve parameterized by $\xi=\beta^2/D$ (Fig. \ref{paramscale}).
From a least squares fit, the scaling relationship was found to be 

\begin{equation}
\label{mscale}
\max(\abdiffm) \propto \left(\frac{\beta^2}{D}\right)^{\frac{1}{6}}
\end{equation}

This scaling as $\hbar^{1/3}$ was also found for a coupled rotor model 
without decoherence \cite{2rotors}.  This result suggests that the 
$\beta^{1/3}$ scaling found here might be generic for systems with more than 
one degree of freedom, and also suggests that the pointwise convergence of 
the quantum probability distribution to the classical distribution 
may occur because of other interacting degrees of freedom (not necessarily 
an external environment).


\subsection{\label{smoothing}Effects of Decoherence vs Smoothing}

The effect of environmental decoherence is analogous to applying a smoothing 
filter to the quantum distributions, but the two effects are not exactly 
the same. 
A small amount of smoothing does not affect the average values of an observable,
whereas decoherence can have an effect by dephasing the 
quantum fluctuation, which are then averaged over (see Fig. \ref{nexpec}).
This effect may be peculiar to systems with only one degree of freedom.

\begin{figure}[tb]
\scalebox{.67}{\includegraphics{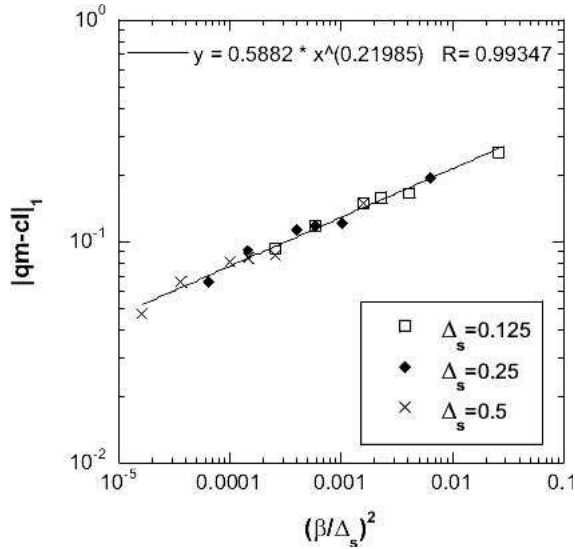}}
\caption{\label{erg_sm} \abdiff for smoothed probability distributions for varying 
 $\sigma_m$ and $\beta$ in the saturation regime ($\tau=20$).}
\end{figure}

Another difference is in the scaling of the QC differences with the parameters.
Figure \ref{erg_sm} shows \abdiff in the saturation regime as a function of 
$\beta$ and $\Delta_s$, where $\Delta_s$ is the half width of the 
triangular filter used to smooth the quantum distribution. 
In the saturation regime, \abdiff was found to depend on the single parameter
 $\beta/\Delta_s$, where it obeys the scaling relation

\begin{equation}
\label{smsc}
\abdiffm=0.58\left(\frac{\beta}{\Delta_s}\right)^{0.44}
\end{equation}

\eqcite{smsc} shows a scaling relationship similar to \eqcite{mscale}.
This may be understood as follows.
Environmental perturbations cause momentum diffusion.  This effect is 
proportional to $\sqrt{D}$, and so $\Delta_s$ should be compared to $ D^2$.  
Hence, in the late time limit, smoothing and decoherence rely on a similar 
composite parameter.  However the power laws are different in the two cases, 
and so the two processes are not entirely equivalent.  


In summary, the environmental perturbations were found to drasticly reduce 
the fine-scale structure in the quantum distributions.  
The measure of the QC differences, \abdiff, was found to initially increase 
with $\tau$ in a form similar to the results in Sec. \ref{chaotic}.  
After reaching a maximum value, \abdiff then decreased exponentially with time.
The decay time was found to be a classical diffusion time, 
and not a decoherence time.  
The maximum QC differences were found to scale as $(\beta^2/D)^{1/6}$, 
where $D$ is the momentum diffusion parameter.

\section{Classical Limit for Hyperion}

\subsection{QC Differences Without Environment}

Having calculated the QC differences for the chaotic rotation of a tidally 
driven satellite, and determined how they scale with the relevant parameters,
we shall now use this information to estimate the magnitude of quantum 
effects on Hyperion.  In particular, we shall assess Zurek's claim 
\cite{zurek,zurek2} that environmental decoherence is needed to ensure 
its classical behavior.  We first examine the magnitude of the QC differences 
for Hyperion if the effect of the environment is ignored.

First we must determine the dimensionless parameter $\beta=\frac{\hbar T}{I_3}$.
Using Hyperion's mean density of $\rho =$ 1.4\ g\ cm$^{-3}$,
and treating it as an ellipsoid with moments of inertia $I_3>I_2>I_1$, 
then $I_{3}=2.1\times 10^{29}$\ kg\ m$^2$. 
Using the value of $\hbar=1.05 \times 10^{-34}$ J\ s, and the orbital period
 $T=1.8\times 10^6$ s, then yields
\begin{equation}
\beta=9.3\times 10^{-58}
\end{equation}

In Sec. IV B it was found that the maximum QC differences in $\avg{J_z}$ 
scale as $\beta^{2/3}$. 
Hence the maximum QC difference in the dimensionless angular momentum
$\avg{J_z}$ for Hyperion should be approximately $5\times 10^{-37}$.
So there should be no observable difference between the quantum and classical 
averages of angular momentum for Hyperion.  

This result contradicts Zurek's claim that, if decoherence is ignored, 
there should be a \textit{break time} of no more than 20 years, beyond 
which the QC differences would become macroscopic.
As was pointed out in the Introduction, if that break time is interpreted 
as the limit of the Ehrenfest regime, then it does not mark the end of the
classical domain. But in \cite{zurek} a \textit{break time} of a similar 
order of magnitude was estimated for the end of the Liouville regime.
Both of those estimates were based on an exponential growth of the 
QC differences that occur in a chaotic state.  Now the deviations from 
Ehrenfest's theorem do, indeed, grow exponentially until they reach the size
of the system, as is needed for Zurek's argument to succeed.  
But, as was shown in Sec. IV(A), the exponential growth of the differences 
between quantum state averages and classical ensemble averages  will 
saturate before those differences reach the size of the system, and the 
saturation value scales with a small power of $\hbar$.  
Hence, for the actual (small) value of $\hbar$, the QC differences in the 
Liouville regime can remain small for all time, and there is no effective 
\textit{break time} for the regime of classicality.

The differences in $\avg{J_z}$ become vanishingly small in the classical limit,
but this does not imply that the full quantum probability distribution 
converges to the classical limit.  We know that the quantum probability 
distribution will not converge to the classical distribution in a pointwise 
fashion.  But we can ask what resolution is needed for a detector to be able 
to discriminate between these two distributions.  
Let us suppose that two probability distributions are practically 
indistinguishable when $\abdiffm<0.01$.  
Using scaling result in Fig. \ref{erg_sm}, we find that a resolution 
$\Delta_s$ of 1 part in $10^{-60}$ rad/s is needed to resolve the two 
probability distributions. 
This suggests that it would be practically impossible to observe the quantum 
effects in the probability distributions, even without invoking environmental decoherence.

\subsection{Environmental Effects on Hyperion}

There are many environmental perturbations that can affect the satellite:
random motion of the particles within the satellite, 
random collisions with interplanetary dust, and random light fluctuations 
from the sun, to name a few.  
We shall consider the random collisions with dust particles as an example.

To do this we treat the interplanetary dust as a dilute gas, and Hyperion as 
a sphere rotating about a fixed axis under the influence of random motion of 
the fluid.  The dimensional momentum diffusion parameter is \cite{coffey}

\begin{equation}
\tilde{D}=16\pi k\mathcal{T}R_h^3\eta
\end{equation}
Using \eqcite{DFeq}, the dimensionless momentum diffusion parameter D is
\begin{equation}
D=\frac{8\pi k\mathcal{T}R_h^3\eta T^3}{I_3^2}
\end{equation}
Here $T$ is the temperature, $k$ is Boltzmann's constant, 
$R_h$ is the radius of Hyperion, and $\eta$ is the kinetic viscosity of 
the dust fluid, which, following \cite{present}, is calculated to be

\begin{equation}
\label{etaeq}
\eta=\frac{nm\bar{v}L}{3\sqrt{2}}
\end{equation}
Here $\bar{v}$ is the rms velocity of the dust particles, $m$ is their mass, 
$n$ is their number density, and $L=1/{n\pi r^2}$ is their mean-free-path, 
and $r$ is the radius of a dust particle.

The properties of interplanetary dust were measured by the Voyager space probes.
The average number density of  
particles near Saturn is $n=4\times 10^{-8}$ m$^{-3}$ \cite{sdust}.  
The average mass of the dust grains is estimated to be $m=10^{-10}$\ g, 
and their radius is about $r=10^{-6}$\ m.  
The temperature in the vicinity of Saturn is about $\mathcal{T}\approx$ 135\ K
 \cite{Caroll}.  

Using \eqcite{etaeq}, treating Hyperion as a sphere of radius 
$R_h=150$\ km, and using $\eta=1.8\times 10^{-6}$ Pa\ s 
for the kinetic viscosity, we estimate the dimensionless angular momentum 
diffusion parameter to be $D\approx 6.4\times 10^{-50}$.  
Even such a small value is sufficient to reduce \abdiff substantially.  
Using \eqcite{mscale}, the order of magnitude of \abdiff for Hyperion is found 
to be $10^{-10}$.  
This implies that the classical and quantum probability distributions will 
agree almost exactly for a large body such as Hyperion.  
Without the influence of the environment, the value of \abdiff 
due to the very fine-scale differences between the quantum and classical 
probability distributions might be of order unity.  But, of course, these 
differences would be impossible to resolve because they exist on such a 
very fine scale.  So the effect of decoherence is to destroy a fine structure 
that would be unobservable anyhow.

\section{Conclusion}

In this paper the regular and chaotic dynamics of a satellite rotating under 
the influence of tidal forces was examined, with application to the motion 
of Hyperion.  Quantum and classical mechanics were compared for both types of 
initial state, and the scaling with $\hbar$ of the quantum-classical (QC)
differences was determined.  
The effect of the environment was modeled, and its effect on the 
QC differences was estimated, so as to determine whether environmental 
decoherence is needed to account for the classical behavior of a macroscopic 
object like Hyperion.
Two measures of the differences between quantum and classical mechanics were 
examined: the QC difference in the average angular momentum,
 $|\avg{J_z}_Q-\avg{J_z}_C|$, and the differences between the probability 
distributions, \abdiff (\eqcite{abdiff}).  

For early times, the QC differences in $\avg{J_z}$ grow in time as $\tau^2$ 
for the non-chaotic state, and as $e^{2.9\tau}$ for the chaotic state.  
At longer times, the QC differences saturate for the chaotic state, but 
oscillate quasi-periodically for the non-chaotic state.  
The magnitude of the QC differences scale as $\beta^2$ (dimensionless $\hbar$) 
at early times, for both the chaotic and non-chaotic states.
This $\beta^2$ scaling persists for all times for the non-chaotic state. 
But the QC differences that occurs in the saturation regime of the 
chaotic state scale as $\beta^{2/3}$.  
A similar scaling has also been observed for a model of two coupled rotors 
 \cite{2rotors}, so this result is not peculiar to the particular model 
studied in this paper.

The value of the dimensionless $\hbar$ for Hyperion is
 $\beta=9.3\times 10^{-54}$, 
for which the $\beta^{2/3}$ scaling relation predicts a maximum value for 
the QC difference in $\avg{J_z}$ to be $5\times 10^{-37}$. 
Therefore, there is no need to invoke environmental decoherence to explain 
the classical behavior of $\avg{J_z}$ for a macroscopic object like Hyperion.


Although the differences between the quantum and classical averages of 
observables becomes very small in the macroscopic limit, this need not be true 
for the differences between quantum and classical probability distributions. 
Indeed, the quantum probability distributions  
do not converge pointwise to the classical probability distributions, 
for either the non-chaotic or the chaotic states. 
A modest amount of smoothing of the quantum distribution reveals that it
is made up of an extremely fine-scale oscillation superimposed upon a smooth
background, and its is that smooth background that converges to the classical
distribution.  Similar behavior has been found for other one-dimensional 
systems \cite{fractal}.  
This smoothing can be regarded as an inevitable consequence of the finite
resolving power of the measuring apparatus. 
Alternatively, it may be impossible to observe the fine structure because 
of environmental decoherence.  At the macroscopic scale of Hyperion, the 
primary effect of decoherence is to destroy a fine structure that is anyhow 
much finer than could ever be resolved by measurement.




When the environment was included, the results were found to follow a 
scaling relationship proposed by \cite{Patyak}: the maximum distance 
between the classical and the quantum probability distributions is 
proportional to $(\beta^2/D)^{1/6}$.  
Here $D$ is the momentum diffusion parameter (see Appendix \ref{diffparam}).  
This suggests that the quantum probability distributions will approach the 
classical distributions pointwise as $\beta \rightarrow 0$,
provided that $D$ is non-zero.  
With environmental perturbations included, the QC differences in the 
probability distributions scaled as $\abdiffm \propto \beta^{1/3}$. 
A similar scaling was also found for two autonomous coupled rotors
 \cite{2rotors}. 
This suggests that pointwise convergence of the quantum probability 
distribution to the classical value may be typical for systems with more 
than one degree of freedom, and the lack of such convergence for systems 
with only one degree of freedom may be pathological.  
The role of the environment, in the model of this paper, is then to cure 
this pathology by supplying more degrees of freedom.

Taking $D$ to be the momentum diffusion parameter for rotation of Hyperion 
due to collisions with the interplanetary dust around Saturn, we find 
(estimated from \eqcite{mscale}) that 
the maximum of \abdiff that Hyperion should exhibit 
should be of order $10^{-10}$. 
Thus decoherence would cause the quantum probability 
distribution to converge to the classical distribution in essentially 
a pointwise fashion.

Coarse graining (due to the finite resolution power of the measurement 
apparatus) will also decrease the QC differences in the probability 
distributions.  In the saturation regime, the measure \abdiff of that 
difference was found to be proportional to $(\beta/\Delta_s)^{0.44}$, 
 where $\Delta_s$ is the width of the smoothing filter. 
This shows that decoherence and smoothing have similar effects.  
But they are not exactly equivalent, since their effects scale with somewhat 
different values of $\beta$.

In conclusion, we find that, for all practical purposes, the quantum theory 
of the chaotic tumbling motion of Hyperion will agree with the classical 
theory, even without taking account of the effect of the environment. 
Decoherence aids in reducing the quantum-classical differences, but it is 
not correct to assert that environmental decoherence is the root cause of 
the appearance of the classical world.

\appendix
\section{Correlated Random Number Generation}

To describe the effect of environmental perturbations, we require a sequence 
of correlated random numbers. Generators for uncorrelated 
random variates are commonly available, but algorithms for generating 
a correlated sequence are not common.  We show here how to generate a 
random sequence having a controlled amount of correlation from a standard 
sequence of independently distributed random numbers.  
Let $\{r_i\}$ be such a sequence, with zero mean and unit variance.
\begin{eqnarray}
\avg{r_i}=0\\
\label{cor2eq}
\avg{r_ir_j}=\delta_{ij}
\end{eqnarray}

To generate a correlated sequence $\{R_i\}$ from the uncorrelated sequence,
we simply form linear combinations, 
\begin{equation}
\label{coreq}
R_{i+1}=cR_i+(1-c)r_{i+1}
\end{equation}
where $c$ is a chosen positive constant $(c<1)$, and $R_1\equiv r_1$.  
It follows from \eqcite{coreq} that

\begin{equation}
\label{coreq2}
R_{i+1}=c^{i-1}r_1+\sum_{m=0}^{i-1}c^m(1-c)r_{i+1-m}
\end{equation}
From this result, we can calculate the degree of correlation in our new 
sequence.
Taking $i>j$, and using \eqcite{coreq2} and \eqcite{cor2eq}, we obtain

\begin{eqnarray}
\label{corr}
\avg{R_iR_j}=\sum_{m=0}^{i-2}\sum_{m'=0}^{j-2}c^{m+m'}(1-c)^2\avg{r_{i+1-m}r_{j+1-m'}}+\nonumber\\\sum_{m=0}^{i-2}c^m(1-c)c^{j-1}\avg{r_1r_{i+1-m}}+\nonumber\\\sum_{m'=0}^{j-2}c^{m'}(1-c)c^{i-1}\avg{r_{j+1-m'}r_1}+c^{i-j-2}\avg{r_1^2}\nonumber\\
\end{eqnarray}
   \eqcite{corr} can be simplified using \eqcite{cor2eq}.
 Performing the resulting
 geometric sums then yields

\begin{equation}
\label{corr2}
\avg{R_iR_j}=\left(\frac{1-c}{1+c}\right)c^{i-1}(c^{1-j}-c^{j-1})
\end{equation}
For $j\gg 1$ and $c<1$, \eqcite{corr2} becomes:

\begin{equation}
\label{corr3}
\avg{R_iR_j}\approx\left(\frac{1-c}{1+c}\right)c^{i-j}
\end{equation}

This discrete sequence must now be converted into a function of time.
Each $R_i$ refers to the correlated random function at time $t_i$. 
Taking the time interval between the random numbers to be $\Delta t$, 
then it is appropriate to define a correlation time $\tau_c=\Delta t/|\ln(c)|$ 
for the correlated random function, for which we have

\begin{equation}
\label{corr4}
\avg{R(0)R(\tau)}\approx\left(\frac{1-c}{1+c}\right)e^{-\tau/\tau_c}
\end{equation}


\section{\label{diffparam} Momentum Diffusion Parameter}

The momentum diffusion parameter ($D$) is needed to calculate the effect of 
the environment on the system \cite{unruh,decondeco,univdeco}.  
In particular, the form of $D$ is needed to show that the scaling result in
 \cite{Patyak} applies also to our model.

Consider the random potential of the form $V=V_0\cos(\phi)$.  
The random torque is then $F=V_0 R(t)\sin(\phi)$.  Here $R(t)$ is a correlated 
random function, as defined in Appendix A.  From this, we can find the momentum diffusion parameter through the relation

\begin{equation}
\label{ddef}
\tilde{D}=\lim_{t\rightarrow 0}\frac{\avg{s(t)^2}}{t}
\end{equation}
The integral of the torque over time yield the angular momentum, hence the 
variance of the angular momentum under this random torque is given by

\begin{eqnarray}
\avg{s^2(t)}=\frac{V_0^2}{2}\int_0^t\!\int_0^tdt'\,dt''\,\avg{R(t')R(t'')}\nonumber\\=\frac{V_0^2}{2}\int_0^t\!\int_0^tdt'\,dt''\,\left(\frac{1-c}{1+c}\right)c^{|t'-t''|/t_c}\nonumber\\\label{tceq}=V_0^2\left(\frac{1-c}{1+c}\right)(t_c^2+t_ct+t_c^2\exp(-t/t_c))
\end{eqnarray}
The quantity $s(t)$ is the standard deviation of the angular momentum for a 
random walk under the influence of \eqcite{ranpot}, and $\sigma$ is the 
standard deviation of the random potential.  
For $t\gg t_c$ we have 
\begin{equation}
\avg{s^2(t)}\approx tt_cV_0^2\left(\frac{1-c}{1+c}\right)
\end{equation}

Using \ref{ddef} and choosing the value $c=1/2$, we obtain momentum 
diffusion constant $\tilde{D}$ to be
\begin{equation}
\label{DFeq}
\tilde{D}=\frac{V_0^2t_c}{3}
\end{equation}

In the body of this paper, we use a dimensionless momentum diffusion 
parameter D.  The relation between these two quantities is

\begin{equation}
\label{DFeq2}
D=\tilde{D}\frac{T^3}{2I_3^2}=\frac{\sigma^2\tau_c}{6}
\end{equation}
where $\tau_c$ is defined to be $\tau_c=t_c/T$, and $\sigma=V_0T^2/I_3$.


\section{\label{patderiv}Derivation of Scaling Parameter}

Pattanayak et al. \cite{Patyak} suggested that at long times the QC differences
in the probability distribution should become a function of the single 
parameter $\xi$, where $\xi=\hbar^a\lambda^bD^c$,
for some powers $a$, $b$ and $ c$. Here $D$ is the diffusion parameter, defined 
in Appendix B, and $\lambda$ is the classical Lyapunov exponent.  

The argument is similar to one presented in \cite{zurek2}. It assumes 
that the differences between quantum and classical mechanics are due to 
the Moyal terms in the equation of motion for the Wigner function.  
These terms have the form

\begin{equation}
\sum_{n=1}^\infty\frac{\hbar^{2n}}{2^{2n}(2n+1)!}\partial_\phi^{2n+1}V(\phi)\partial^{2n+1}_p\rho^w
\end{equation}
where $\rho^w$ is the Wigner function for the state.  
For a chaotic system, the phase space distribution will develop very fine  
structures as it fills the accessible phase space, with the rate at which 
these fine structures develop being governed by the Lyapunov exponent $\lambda$.  
Since these terms depend on $\partial^{2n+1}_p \rho^w$, they will become 
larger as time progresses and the fine structure grows.  
However the inclusion of environmental perturbations on system causes 
diffusion, which will limit the growth of the fine structure.  
When these effects balance each other, the fine structure is expected to 
have an equilibrium scale \cite{Patyak,zurek2} given by

\begin{equation}
\diff{\rho^w}{p}\approx \sqrt{\frac{\lambda}{2D}}
\label{metascale}
\end{equation}
If this equilibrium momentum scale is sufficiently large compared to $\hbar$, 
then only the first order Moyal term should be significant.  
Under this assumption, the QC differences should be a function of this 
Moyal term, given by
\begin{equation}
\xi=\hbar^{2}\lambda^{5/2}D^{-(3/2)}\diffm{V(\phi)}{\phi}{3}
\label{xieq}
\end{equation}

It is argued \cite{zurek2} that the characteristic scale on which $\rho^w$ 
 varies is  $\delta \phi\propto \sqrt{D/\lambda}$.  
The characteristic variation of $\partial^3_\phi V(\phi)$ can be found by 
a Taylor expansion of  $\partial^3_\phi V(\phi) \propto \sin(\phi)$ about 
an arbitrary point $\phi_0=0$. 
Averaging the result over $\phi_0=0$ then yields


\begin{equation}
\diffm{V(\phi)}{\phi}{3}\propto \delta \phi\approx\sqrt{\frac{D}{\lambda}}
\label{vvar}
\end{equation}
Inserting \eqcite{vvar} into \eqcite{xieq} yields

\begin{equation}
\xi\propto \frac{\hbar^2}{D}
\end{equation}

Thus, once the growth fine scale structure of the probability distribution 
reaches equilibrium with diffusion, the QC differences in the probability 
distribution should be a function of $\hbar^2D^{-1}$.
The results in Sec. \ref{decoh} confirm this conclusion.

\acknowledgments
We would like to thank J. Emerson and B. C. Sanders for  
many helpful suggestions.
This work was supported by the Natural Sciences and Engineering Research 
Council of Canada.


\end{document}